\documentclass{eas}
\usepackage{graphicx}
\begin{document}

\title{Observations of Protostellar Outflow Feedback in Clustered Star Formation} 
\author{Fumitaka Nakamura}\address{National Astronomical Observatory of Japan,
2-21-1 Osawa, Mitaka, 181-8588, Japan}

\begin{abstract}
We discuss the role of protostellar outflow feedback in clustered star formation
using the observational data of recent molecular outflow surveys toward nearby cluster-forming clumps.
We found that for almost all clumps, the outflow momentum injection
rate is significantly larger than the turbulence dissipation rate. Therefore, the
outflow feedback is likely to maintain supersonic turbulence in the clumps.
For less massive clumps such as B59, L1551, and L1641N, the outflow kinetic energy is
comparable to the clump gravitational energy. In such clumps, the outflow feedback probably affects 
significantly the clump dynamics. On the other hand, for clumps with masses 
larger than about 200 M$_\odot$, the outflow kinetic energy is significantly smaller than 
the clump gravitational energy.
Since the majority of stars form in such clumps, we conclude that 
outflow feedback cannot destroy the whole parent clump.
These characteristics of the outflow feedback support the scenario of slow star formation.
\end{abstract}
\maketitle

\runningtitle{Nakamura \etal: Observations of Stellar Feedback}

\vspace*{-0.3cm}
\section{Introduction}

Most stars form in clustered environments (\cite{lada10}). 
Radio and infrared observations have revealed that star clusters form in parsec-scale dense
molecular clumps with masses of the order of 10$^2- 10^3$ M$_\odot$.
In such clumps, stellar feedback from forming stars shapes their internal structure and affects 
the formation of next-generation stars.
Therefore, understanding the role of stellar feedback in the process of cluster formation is a central problem of 
star formation study.

There are several types of stellar feedback such as outflows, winds, and radiation (\cite{bally10}, \cite{krumholz14}). 
Among them, some theoretical studies suggest that the radiation pressure from high-mass stars is the most dominant 
feedback mechanism in the presence of high-mass stars. However, in the process of high-mass star formation, 
the outflow feedback from low-mass stars may play a dominant role in regulating star formation 
before high-mass stars are formed.
In other words, the outflow feedback can control the formation of high-mass stars by injecting momentum and energy 
in the surrounding gas. In addition, there are many cluster-forming regions where no high-mass stars form. 
In such regions, outflow feedback should play a dominant role in regulating star formation. 
In this contribution, we focus on the protostellar outflow feedback and discuss its role on the clump dynamics on the basis of observations.

Protostellar outflows have been shown theoretically to be capable of maintaining supersonic turbulence in
cluster-forming clumps and keeping the star formation rate per free-fall time as low as a few percent (\cite{nakamura07}). 
However, its exact role in clustered star formation remains controversial.
Two main scenarios have been proposed for the role of outflow feedback in clustered star formation. 
In the first scenario, the outflow feedback is envisioned to destroy the cluster-forming clump as a whole, 
which terminates further star formation. In this case, star formation should be rapid and brief (e.g., \cite{maclow04}). 
On the other hand, in the second scenario, the outflow feedback is envisioned to play the role of maintaining 
the internal turbulent motions. In this scenario, star formation should be slow
and can last for several free-fall times or longer (e.g., \cite{tan06}).
Below we constrain these theoretical models using the observational results of the protostellar outflow feedback.

\vspace*{-0.3cm}
\section{Data: Protostellar Outflows}

\begin{table}[htb]
  \begin{tabular}{lcccl} \hline
    Clumps       & distance & Mass        & Radiius  & References \\ 
                 & (pc)     & (M$_\odot$) & (pc)     &  \\ \hline 
    B59          & 130 & 30  & 0.3  & \cite{duarte12} \\
    L1551        & 140 & 110 & 1.0  & \cite{sto06} \\
    L1641N       & 400 & 210 & 0.55 & \cite{nakamura12} \\ 
    Serpens Main & 415 & 535 & 0.73 & \cite{sugitani10} \\
    Serpens South& 415 & 232 & 0.2  & \cite{nakamura11b} \\
    $\rho$ Oph   & 125 & 883 & 0.8  & \cite{nakamura11a} \\
    IC 348       & 250 & 620 & 0.9  & \cite{arce10} \\
    NGC 1333     & 250 & 1100& 2.0  & \cite{arce10} \\ 
    NGC 2264-C   & 800 & 2300& 0.7  & \cite{maury09} \\ \hline
  \end{tabular}
  \caption{Nearby cluster-forming clumps. We note that the clump masses are estimated by $^{13}$CO data except
for Serpens South, for which the Herschel data are used because of severe absorption of the $^{13}$CO emission.} 
  \label{tab:1}
\end{table}

The outflowing gas from protostars accelerates entrained gas 
to large velocities. Such components are observed as molecular 
outflows and such molecular outflows are important to inject 
momentum in the surrounding gas.
The CO lines such as the $J=1-0$, $J=2-1$, and $J=3-2$ transitions  
are excellent tracers to measure the parameters of molecular outflows. 
In Table \ref{tab:1}, we summarize some properties of nearby cluster-forming clumps 
toward which CO outflow surveys were carried out.
In the next section, we discuss how CO outflows affect the parent clumps
using the observational results presented in the references listed in Table \ref{tab:1}.

Outflow feedback has two different roles. One is negative effect, slowing 
down or terminating star formation by injecting momentum and energy. 
In this case, the mean densities of regions decrease by feedback. 
Another role is positive effect, triggering future star formation 
by dynamically compressing the surroundings. 
In this case, the local densities increase by compression.
For outflow feedback, the negative effect is likely to be more important 
than positive one because the outflow feedback happens inside the star-forming 
regions, dispersing gas outwards. 
Therefore, in the following we focus on the negative effect of the outflow feedback.
In the next section, we address the following two important questions of 
star formation using the observational data.
\begin{itemize}
\item[(1)] Can protostellar outflow feedback maintain supersonic turbulence in clustered environment?
\item[(2)] Can protostellar outflow feedback directly destroy the parent clumps?
\end{itemize}

\begin{table}[htb]
\begin{center}
  \begin{tabular}{lll} \hline
Model   & (1) & (2)   \\ \hline 
Rapid Star Formation    & Yes/No & Yes   \\
Slow Star Formation     & Yes    & No    \\ \hline
  \end{tabular}
  \caption{Rapid vs. slow star formation models.}
  \label{tab:2}
\end{center}
  \end{table}

The two models of star formation give the different answers to these questions.
In Table \ref{tab:2}, we summarize the predicted answers of these two questions 
for the two star formation models.

\vspace*{-0.3cm}
\section{Results}

\begin{table}[htb]
\begin{center}
  \begin{tabular}{lccclll} \hline
    Clumps       & $\alpha$ & $\sigma$  & $dP_{\rm turb}/dt$  & $dP_{\rm out}/dt$ & $E_{\rm out}$ & $E_{\rm grav}$  \\ \hline 
    B59          & 1.1 & 0.4  & 1.0 $ \times 10^{-5}$ & 8.5 $ \times 10^{-5}$ & 4  & 13  \\
    L1551        & 1.3 & 0.45 & 1.8 $ \times 10^{-5}$ & 6.3 $ \times 10^{-4}$ & 130& 52  \\
    L1641N       & 1.0 & 0.74 & 1.3 $ \times 10^{-4}$ & 1.3 $ \times 10^{-3}$ & 273& 581 \\ 
    Serpens Main & 0.7 & 0.85 & 3.4 $ \times 10^{-4}$ & 2.5 $ \times 10^{-3}$ & 445& 1686\\
    Serpens South& 0.2 & 0.53 & 2.1 $ \times 10^{-4}$ & 6.5 $ \times 10^{-4}$ & 165& 1157\\
    $\rho$ Oph   & 0.2 & 0.64 & 2.9 $ \times 10^{-4}$ & 1.2 $ \times 10^{-4}$ & 61 & 4191\\
    IC 348       & 0.6 & 0.76 & 2.5 $ \times 10^{-4}$ & 4.7 $ \times 10^{-4}$ & 26 & 1837\\
    NGC 1333     & 1.1 & 0.93 & 3.0 $ \times 10^{-4}$ & 1.1 $ \times 10^{-3}$ & 119& 2602\\ 
    NGC 2264-C   & 0.6 & 1.7  & 5.5 $ \times 10^{-4}$ & 1.7 $ \times 10^{-3}$ & 50  &  27652 \\ \hline
  \end{tabular}
  \caption{Some physical parameters of the outflow feedback for nearby cluster-forming clumps. 
$\alpha$ is the virial ratio, $\sigma$ is 1D velocity dispersion (km s$^{-1}$),
$dP_{\rm turb}/dt$ is the dissipation rate of turbulence ($M_\odot$ km s$^{-1}$ yr$^{-1}$),
$dP_{\rm out}/dt$ is the outflow momentum injection rate ($M_\odot$ km s$^{-1}$ yr$^{-1}$), 
$E_{\rm out}$ is the outflow kinetic energy ($M_\odot$ km$^{2}$ s$^{-2}$),
$E_{\rm grav}$ is the clump gravitational energy ($M_\odot$ km$^{2}$ s$^{-2}$).}
\end{center}
\label{tab:3}
\end{table}

First, using the observational data of the outflow surveys listed in Table \ref{tab:1}, 
we calculated some physical parameters of the outflow feedback in Table \ref{tab:3}.

Here, we assumed that outflow gas is optically-thin and that the inclination angles of all the identified outflows are fixed to be 
the mean value of 57.3 degree. 
However, low-velocity wings sometimes become optically thick. Therefore, the assumption of the optically-thin leads to underestimation of 
the outflow mass by a factor of 10 or more.  However, we note that we may underestimate momentum and energy only by a factor of a few 
because the optically-thick gas have low-velocity, and have less momentum and energy 
than high-velocity components.

\begin{figure}
\begin{center}
\includegraphics[width=10cm]{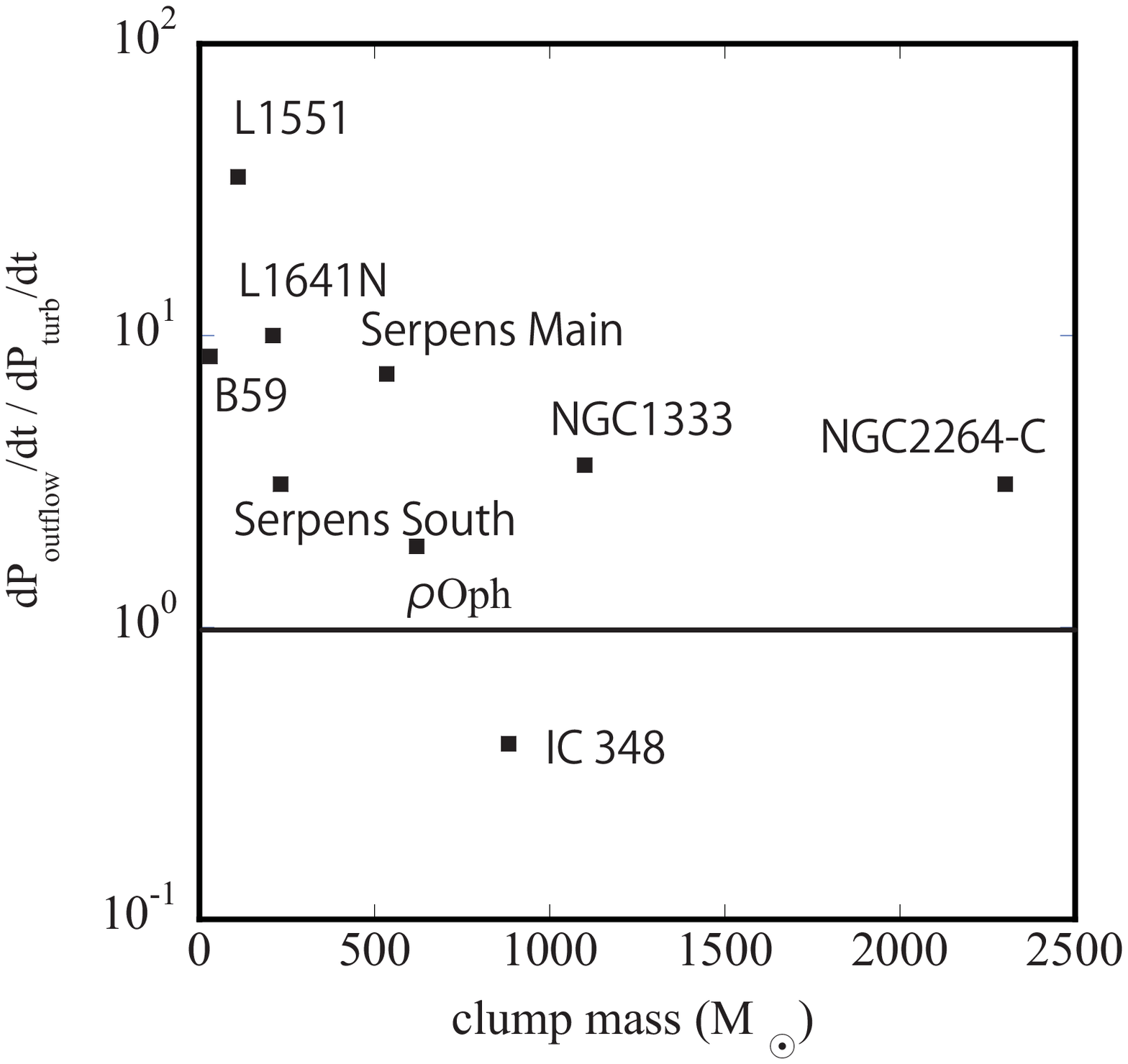}
\caption{Ratios of the momentum injection to the turbulence dissipation rates
of nearby cluster-forming clumps as a function of clump mass. 
Our result seems to be inconsistent with that of Maury et al. (2006) for
NGC 2264-C.
The difference comes from the fact that they use the energy dissipation rate, 
instead of the momentum dissipation rate. 
In addition, they omitted the numerical factor of 0.21 to estimate the dissipation
rate of turbulence (see \cite{nakamura14}).}
\end{center}
\label{fig:1}
\end{figure}

\vspace*{-0.3cm}
\subsection{Can outflow feedback maintain supersonic turbulence?}
	
To answer this question, we compare the turbulence dissipation rate ($dP_{\rm turb}/dt$) and outflow 
momentum injection rates ($dP_{\rm out}/dt$) towards target clumps. 
In previous studies, the energy is used for the analysis. 
However, because the outflow feedback is the momentum feedback (\cite{krumholz14}), 
we use momentum instead of energy.
The dissipation rate of turbulence is defined as the total cloud momentum divided by 
turbulence crossing time ($dP_{\rm turb}/dt=0.21 M_{\rm cl} \sigma_{\rm 3D} /t_{\rm diss}$). 
The numerical factor of 0.21 is determined from comparison with turbulence simulations. 
The outflow momentum injection rate is defined by the outflow momentum 
divided by the outflow dynamical time.

In Figure 1, we show the ratio between momentum injection rate and dissipation rate as a function of clump mass. 
The momentum injection rates are larger than the dissipation rates for all clumps except IC 348. 
Thus, we conclude that the outflow feedback has enough momentum to maintain supersonic turbulence in these clumps.

\begin{figure}
\begin{center}
\includegraphics[width=10cm]{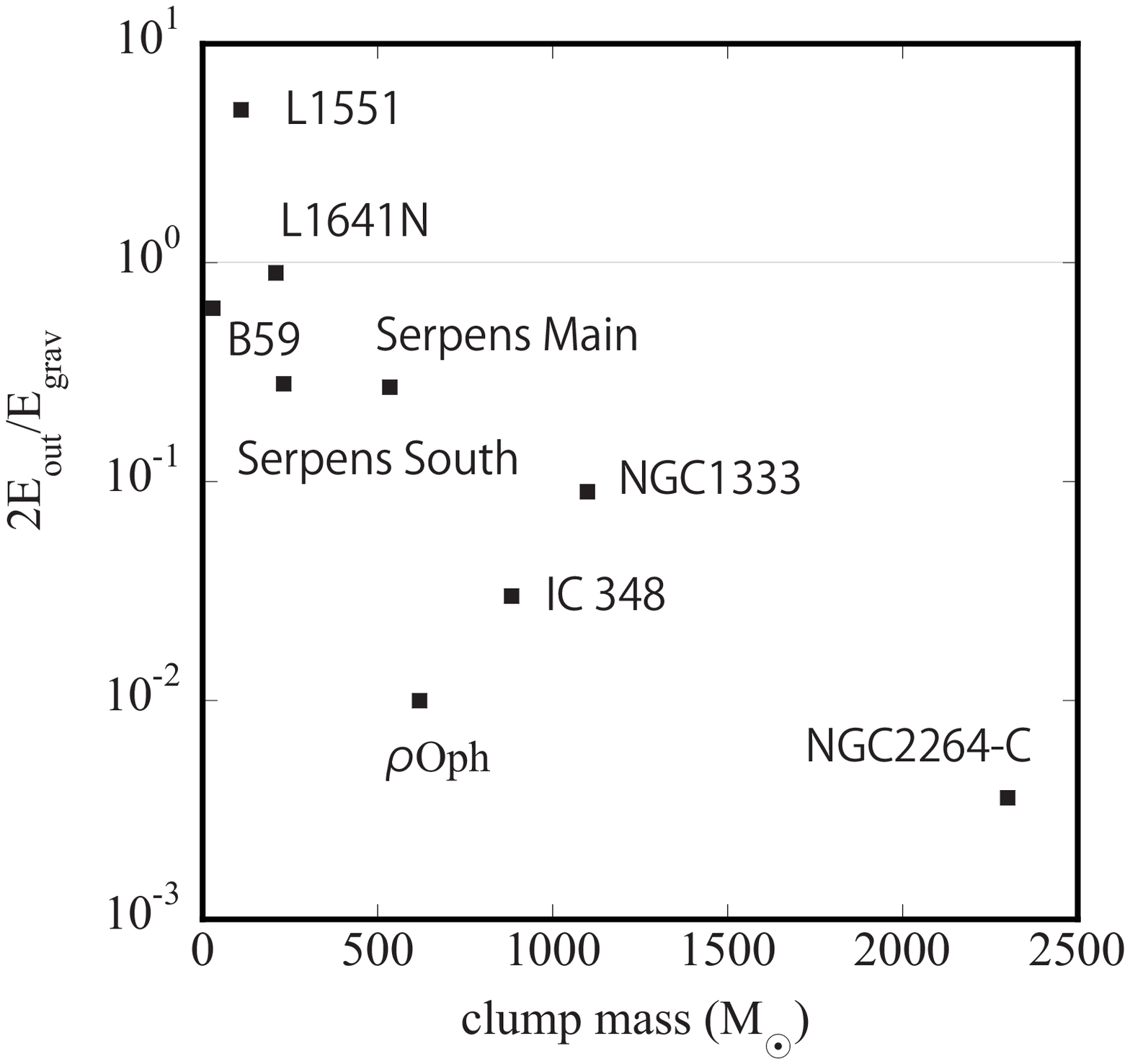}
\caption{Ratios of twice the outflow kinetic energy $2E_{\rm out}$ to the gravitational
binding energy of the clump $E_{\rm grav}$ for nearby cluster-forming clumps as a
function of clump mass.}
\end{center}
\label{fig:2}
\vspace*{-0.5cm}
\end{figure}

\vspace*{-0.3cm}
\subsection{Can outflow feedback directly destroy the parent clump?}

To answer this question, we adopt the virial theorem, $d^2I/dt^2= 2E_{\rm cl} - E_{\rm grav}$. 
First, we list the virial ratios ($\alpha = 2E_{\rm cl}/E_{\rm grav}$) of the clumps
in Table \ref{tab:3}. Almost all clumps are close to virial equilibrium 
within a factor of a few.  
In Figure 2, we show the ratios of $2 E_{\rm out}$ and $E_{\rm grav}$ as a function of clump mass. 
Figure 2 indicates that for less massive clumps (B59, L1551, L1641N, $<200$ M$_\odot$), 
the outflow energy is comparable to the gravitational energy. 
The dynamics of such clumps may be significantly influenced by the outflow feedback.
On the other hand, for more massive clumps ($>$ 200 M$_\odot$), the outflow kinetic energy 
appears to be significantly smaller than the gravitational energy. 
In other words, the outflow feedback may not directly destroy the whole clumps. 
But, a fraction of gas may be dispersed by outflows. This gentle ejection of gas 
may lead to clump dispersal eventually.

\vspace*{-0.5cm}
\section{Slow vs. Rapid Star Formation}

From the results presented in the previous section, we found that the outflow feedback
can maintain supersonic turbulence in the nearby cluster-forming clumps.
In addition, the outflow kinetic energy is significantly smaller than
the clump gravitational energy except for the three least
massive clumps, B59, L1551, and L1641N. Therefore, we
conclude that the outflow feedback is not enough to disperse
the whole clump at least for the clumps with masses greater than 200 M$_\odot$. 
Since the majority of stars form in such clumps, 
we conclude that the observations of the outflow feedback 
support the slow star formation.


\end{document}